\documentclass[aps,prd,twocolumn,showpacs]{revtex4} 

\usepackage{graphicx}

\font\FermiSmallfont=cmssq8 scaled 1200

\def\LANLppthead#1{
\null 
\begin{center}\vskip -1.0truein{\hbox to 7.5truein {
\hfill
\vbox to 1in {\vfill \FermiSmallfont
              \hbox{#1}
              \vfill}
}}\vskip-0.0truein\end{center}}

\begin{document}

\title{Weak Interaction Rate Coulomb Corrections in Big Bang Nucleosynthesis}
\author{Christel J. Smith and George M.\ Fuller}
\affiliation{Department of Physics, University of California, San Diego, La Jolla, CA
92093-0319}

\date{\today}

\begin{abstract}

We have applied a fully relativistic Coulomb wave correction to the weak reactions in the full Kawano/Wagoner Big Bang Nucleosynthesis (BBN) code.  We have also added the zero temperature radiative correction.  We find that using this higher accuracy Coulomb correction results in good agreement with previous work, giving only a modest $\sim 0.04\%$ increase in helium mass fraction over correction prescriptions applied previously in BBN calculations.  We have calculated the effect of these corrections on other light element abundance yields in BBN and we have studied these yields as functions of electron neutrino lepton number.  This has allowed insights into the role of the Coulomb correction in the setting of the neutron-to-proton ratio during the BBN epoch.  We find that the lepton capture processes' contributions to this ratio are only second order in the Coulomb correction.

\end{abstract}
\maketitle

\section{Introduction}

The study of Big Bang Nucleosynthesis (BBN) has been and is a powerful tool for testing cosmological models and constraining the fundamental parameters of the universe.  Since primordial nucleosynthesis occurs relatively soon after the Big Bang ($\sim 1\,$s), BBN provides  one of the best windows into the physics of the early universe.  

Before the launch of the Wilkinson Microwave Anisotropy Probe (WMAP), BBN predictions along with direct observation of the primordial element abundances were used to constrain the baryon-to-photon ratio, $\eta$.  The independent high precision determination of $\eta$ from the ratio of the acoustic peak amplitudes in the cosmic microwave background (CMB) from WMAP \cite{WMAP, WMAP1,3yrwmap} allows us now to use BBN to constrain other unknowns in the early universe and physics beyond the Standard Model.  

The CMB measurement of $\eta$ increases in precision with accumulating WMAP data and the future Planck mission promises even higher precision, with a projected $\sim 1\%$ accuracy in $\eta$ \cite{3yrwmap, bond}.  The measurement of primordial deuterium also shows promise for higher accuracy determination as more QSO lines of sight become available \cite{Tytler, Omeara} .  This fuels the motivation to further refine the calculation of predicted primordial element abundances.  For this reason we have analyzed the effect of adding the full relativistic Coulomb wave correction factor (relativistic Coulomb barrier factor) to the weak reaction rates in the BBN calculation.

Among many issues, a key piece of physics that sets the stage for primordial element nucleosynthesis is the evolution of the neutron-to-proton ratio, $n/p$.  The $n/p$ ratio is critical in determining the synthesis of the primordial elements because it sets the number of neutrons available to build nuclei.  

The neutron-to-proton ratio is effectively determined by the competition between the charge-changing weak interaction rates and the expansion rate of the universe.  Listed below are the weak reactions which interconvert neutrons and protons:
\begin{eqnarray}
& \nu_e+n\rightleftharpoons p+e^-, \label{nuenn} \\
& \bar\nu_e+p\rightleftharpoons n+e^+, \label{nuebarpp} \\
& n \rightleftharpoons p+e^-+\bar\nu_e. \label{ndecayy}
\end{eqnarray}
The corresponding rates for these weak reactions are denoted by $\lambda_{\nu_en}$ and $\lambda_{e^-p}$, $\lambda_{\bar\nu_ep}$ and $\lambda_{e^+n}$, $\lambda_{n_{\rm decay}}$ and $\lambda_{pe^-\bar\nu_e}$ for the forward and reverse reactions in Eq.~(\ref{nuenn}), Eq.~(\ref{nuebarpp}), and Eq.~(\ref{ndecayy}), respectively.  Defining $\Lambda_n=\lambda_{\nu_en} + \lambda_{e^+n} +\lambda_{n_{\rm decay}}$ and $\Lambda_{\rm tot} = \Lambda_n + \lambda_{\bar\nu_e} + \lambda_{e^-p}+\lambda_{pe^-\bar\nu_e}$, and defining the neutron-to-proton ratio to be $n/p$, we can show that in the early universe
\begin{equation}
{d\over{dt}}\left({n\over{p}}\right) = (1+n/p)^2\left({\Lambda_{\rm tot}\over{1+n/p}} - \Lambda_n\right),
\end{equation}
where $t$ is the Freidmann-Lemaitre-Robertson-Walker timelike coordinate \cite{abfw}.  Note that the net number of electrons minus positrons per baryon is $Y_e \equiv (n_{e^-} - n_{e^+})/{n_b} = (1+n/p)^{-1}$.  At high temperatures, $T \gg 1$ MeV, the weak reaction rates are fast compared to the expansion rate of the universe, steady state equilibrium (${d\over{dt}}(n/p)=0$) is a good approximation, and the neutron-to-proton ratio is given by \cite{abfw, ourcode}
\begin{equation}
{{n}\over{p}}  = {{\lambda_{\bar\nu_ep}+\lambda_{e^-p}+\lambda_{pe^-\bar\nu_e}}\over{\lambda_{\nu_en}+\lambda_{e^+n}+\lambda_{n\ {\rm decay}}}}.
\label{ntoppp}
\end{equation}

\begin{figure}
\includegraphics[width=2.5in,angle=270]{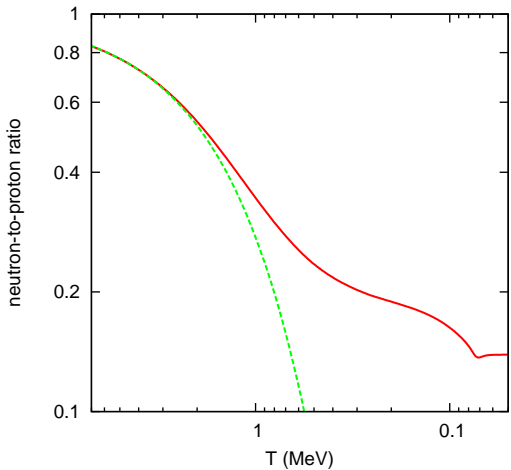}
\caption{Neutron-to-proton ratio as a function of temperature. The full standard BBN zero lepton number case is given by the solid line.  The dashed line is the neutron-to-proton ratio as calculated with an enforced assumption of steady state equilibrium, $i.e.$, Eq.~(\ref{chemnp}).}
\label{figurea}
\end{figure}

The relatively slow rates at high temperatures for both free neutron decay and the corresponding reverse three body reaction allow the steady state equilibrium neutron-to-proton ratio to be approximated as
\begin{equation}
\label{ntoppp}
{{n}\over{p}}  \approx  {{\lambda_{\bar\nu_ep}+\lambda_{e^-p}}\over {\lambda_{\nu_en}+\lambda_{e^+n}}}.
\end{equation}
If the neutrinos have thermal, Fermi-Dirac energy distribution functions, this can be written by
\begin{equation}
\label{thernalnp}
{{n}\over{p}} \approx {{\left(\lambda_{e^-p}/\lambda_{e^+n}\right)+e^{-\eta_{\nu_e}+\eta_e-\xi}}\over{\left(\lambda_{e^-p}/\lambda_{e^+n}\right) e^{\eta_{\nu_e}-\eta_e+\xi}+1}},
\end{equation}
where $\eta_{\nu_e}=\mu_{\nu_e}/T$ is the electron neutrino degeneracy parameter, $\eta_e=\mu_e/T$ is the electron degeneracy parameter, with $\mu_{\nu_e}$ and $\mu_e$ the electron neutrino and electron chemical potentials, respectively, and $\xi$ is the neutron-proton mass difference divided by temperature, $\xi= \delta{m_{np}}/T$ with $\delta m_{np} = m_nc^2-m_pc^2\approx 1.293$\,MeV, where we take the Boltzmann constant to be $k_{\rm B}=1$ \cite{abfw}.

At high temperatures the weak reactions can be fast enough to maintain chemical equilibrium.  In chemical equilibrium, the chemical potentials satisfy the Saha equation, $\mu_{\nu_e} + \mu_n = \mu_{e^-} + \mu_p$.  When chemical equilibrium is maintained the neutron-to-proton ratio will be
\begin{equation}
\label{chemnp}
{{n}\over{p}} \approx e^{\left({\mu_e-\mu_{\nu_e}-\delta m_{np}}\right)/{T}}.
\end{equation}
This result can be obtained directly from the ratio of the appropriate Fermi-Dirac distribution functions or, alternatively and equivalently so long as all reactants have a Fermi-Dirac form for their energy spectra, from evaluation of the rates in Eq.~(\ref{thernalnp}) \cite{abfw}.

As the universe cools and expands, the rates of the weak reactions become slow compared to the expansion rate of the universe.  At this point chemical equilibrium can no longer be maintained and a period known as the \lq\lq weak freeze out\rq\rq\ occurs as the neutron-to-proton ratio pulls away from its equilibrium value.  For a faster expansion rate of the universe the weak reaction rates become comparatively slow earlier and the neutron-to-proton ratio falls out of equilibrium at higher temperatures, yielding a relatively larger $n/p$ value.  Likewise, for a slower expansion rate, the weak reactions can maintain chemical equilibrium longer and the neutron-to-proton ratio consequently would be lower.  The expansion rate of the universe is set by the local total energy density through the Freidman equation. 

Figure\,\ref{figurea} shows the neutron-to-proton ratio as a function of temperature for a standard Big Bang scenario with zero lepton numbers, $i.e.$, $\mu_{\nu_e} = \mu_{\bar\nu_e} =\mu_e = 0$.  This figure shows both the actual $n/p$ ratio and the approximation to this with an enforced chemical equilibrium condition.  Obviously, these agree for high temperature but diverge once the weak reaction rates become slow compared to the expansion rate of the universe.  Note that the actual $n/p$ ratio becomes constant once nearly all free neutrons are incorporated into alpha particles at $T < 100$\, keV.  This figure shows the $n/p$ ratio only for free neutrons and neutrons bound within alpha particles, and neglects the neutrons bound in $^2$H, $^3$H, $^3$He, and nuclei heavier than $^4$He, accounting for the small divit in the upper curve near $T=$80 keV.

Primordial element abundance yields are calculated by a BBN code that time evolves the temperature and expansion rate of the universe along with the nuclear and weak reactions rates.  We have used a modified version of the Kawano/Wagoner BBN code \cite{kawano, kawano1, wag69, wfh} in order investigate the effect of integrating the relativistic Coulomb barrier factor in the appropriate weak reaction rates.  In section II we discuss the calculation of the weak reaction rates.  In section III we discuss the relativistic Coulomb correction employed here and Coulomb correction prescriptions studied previously.  In section IV we present results and give a discussion and in section V we give conclusions.

\section{The Weak Reaction Rates}

We calculate the individual weak interaction rates with the following phase space factor forms and with a common matrix element which is proportional to the inverse of an effective ft-value $\langle ft \rangle $ \cite{abfw, dicus, FFNI, FFNII, FFNIII, FFNIV,ourcode}:
\begin{widetext}
\begin{equation}
\lambda_{e^-p}  \approx {{ \ln{2}}\over{\langle ft\rangle { \left( m_ec^2 \right)}^5  }}
\int_{\delta m_{np}}^\infty {{F \left[Z,E_e\right] (E_e-\delta m_{np})^2 E_e\left(E_e^2 -m_ec^2\right)^{1/2} }\left[S_{e^-}\right] \left[ 1-{{S}}_{\nu_e} \right] dE_e },\label{genep} 
\end{equation}

\begin{equation}
\lambda_{\bar\nu_e p} \approx {{ \ln{2}}\over{\langle ft\rangle {\left(m_ec^2 \right)}^5 }}
\int_{m_ec^2}^\infty {{(E_e+\delta m_{np})^2 E_e( E_e^2  -m_ec^2)^{1/2}}\left[S_{\bar\nu_e} \right] \left[1-S_{e^+}\right] dE_e}, \label{nuonprate} 
\end{equation}

\begin{equation}
\lambda_{e^+n} \approx {{ \ln{2}}\over{\langle ft\rangle {\left(m_ec^2 \right)}^5 }}
\int_{m_ec^2}^\infty (E_e+\delta m_{np})^2 E_e( E_e^2  -m_ec^2)^{1/2} \left[S_{e^+}\right] \left[1-S_{\bar\nu_e}\right] dE_e, \label{eonnrate} 
\end{equation}

\begin{equation}
\lambda_{\nu_e n} \approx {{ \ln{2}}\over{\langle ft\rangle {\left(m_ec^2 \right)}^5 }}
\int_{\delta m_{np}}^\infty{{F \left[Z,E_e\right] (E_e-\delta m_{np})^2 E_e ( E_e^2 -m_ec^2)^{1/2}}\left[S_{\nu_e}\right] \left[1-S_{e^-}\right] dE_e}, \label{nueonrate}
\end{equation}

\begin {equation}
\lambda_{n_{\rm decay}} \approx {{ \ln{2}}\over{\langle ft\rangle {\left(m_ec^2 \right)}^5 }}
\int_{m_ec^2}^{\delta m_{np}} {F \left[Z,E_e\right](\delta m_{np}-E_e)^2 E_e \left( E_e^2 -m_ec^2\right)^{1/2}} \left[1-S_{\bar\nu_e}\right] \left[1-S_{e^-} \right] dE_e , \label{ndecayrate}
\end{equation}

\begin{equation}
\lambda_{pe^-\bar\nu_e} \approx {{ \ln{2}}\over{\langle ft\rangle {\left(m_ec^2 \right)}^5 }}
\int_{m_ec^2}^{\delta m_{np}} {F \left[Z,E_e \right] (\delta m_{np}-E_e)^2 E_e \left( E_e^2 -m_ec^2\right)^{1/2}}  \left[S_{\bar\nu_e}\right] \left[S_{e^-} \right] dE_e ,\label{revrate}
\end{equation}

\end{widetext}

where $E_e$ is the total electron or positron energy as appropriate, $m_ec^2$ is the electron rest mass, and $F\left[Z,E_e\right]$ is the Coulomb correction Fermi factor which will be discussed in detail below.  Note that the nuclear charge relevant here is $Z=1$.  $S_{e^{-/+}}$ and $S_{\nu_e/\bar\nu_e}$ are the phase space occupation probabilities for electrons/positrons and neutrinos/antineutrinos, respectively.  For neutrinos and electrons with energy distributions with the expected thermal form, the occupation probabilities are
\begin{equation}
S_{\nu_e} = {1\over { e^{E_{\nu}/{T_\nu} - \eta_{\nu_e}} +1}},
\label{nuocc}
\end{equation}
\begin{equation}
S_{\bar\nu_e} = {1 \over{ e^{E_{\nu}/{T_\nu} - \eta_{\bar\nu_e}} +1}},
\label{nubarocc}
\end{equation}
\begin{equation}
S_e = {1 \over{ e^{E_e/{T}} +1}},
\label{elecocc}
\end{equation}
where $T_\nu$ is the neutrino temperature parameter, $\eta_\nu$ is the neutrino degeneracy parameter (the ratio of chemical potential to temperature), and $E_\nu$ is the appropriate neutrino or antineutrino energy.

We take
\begin{equation}
{\ln 2\over{\langle ft\rangle}} = {c\,(m_ec^2)^5\over{\hbar c}}\cdot\delta\cdot{G_F^2 |C_V|^2\left(1+3|C_A/C_V|^2\right)\over{2\pi^3}},
\label{logft}
\end{equation}
where $G_F \approx 1.166\times 10^{-11}$\,MeV$^{-2}$ is the Fermi constant, $C_V$ and $C_A$ are the vector and axial vector coupling constants, respectively, and we have taken the absolute squares of the Fermi and Gamow-Teller matrix elements for the free nucleons to be $|M_F|^2 = 1$ and $|M_{GT}|^2 = 3$, respectively.  Here $\delta$ is a factor which includes both Coulomb and other (\lq\lq radiative correction\rq\rq) effects which amount to a few percent change in the effective $ft$-value, $\langle ft \rangle$.  

Of course, $C_V$ and $C_A$ are coupling constants that are renormalized by the particular strong interaction environment characterizing free neutrons and protons. (Absent strong interactions $C_V=C_A=1$.)  Given that these are {\it a priori} unknowns, as is $\delta$, we follow the standard procedure \cite{wfh}: we take the free neutron decay rate as the product of Eq.~(\ref{logft}) and the phase space factor in Eq.~({\ref{ndecayrate}) (with $S_{\bar\nu_e}=S_{e^-}=0$) and we then set this equal to the inverse of the laboratory-measured free neutron lifetime, $\tau_n$.  The world-average of the laboratory measurements is $\tau_n = 887.7$ seconds \cite{pdg}.

Note that changing the prescription for the Coulomb correction factor $F[Z,E_e]$ in Eq.~(\ref{ndecayrate}) will have the effect of renormalizing the effective free nucleon weak interaction matrix elements ($i.e.,$ renormalizing $\langle ft\rangle$) for a given $\tau_n$.  As we will see below, this renormalization will be the dominant component of the Coulomb correction alteration in, $e.g.$, the $^4$He BBN yield.

The rates for all the individual weak reactions are shown as functions of temperature in Fig.~\ref{figure1}.  At high temperatures the forward and reverse rates of the lepton capture reactions in Eq.~(\ref{nuenn})  and Eq.~(\ref{nuebarpp}) dominate the neutron-proton inter-conversion process.  Note that the rates for the forward process in Eq.~(\ref{nuebarpp}) and the reverse process in Eq.~(\ref{nuenn}) are affected by the threshold, $\delta m_{np}+m_ec^2$.  At lower temperatures this threshold makes these rates relatively slower than the rates for the lepton capture channels without this threshold, $i.e.$, the forward process in Eq.~(\ref{nuenn}) and the reverse process in Eq.~(\ref{nuebarpp}).

\begin{figure}
\includegraphics[width=2.5in,angle=270]{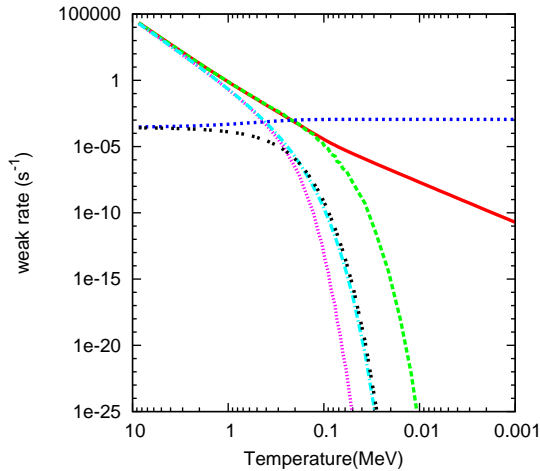}
\caption{All six weak reaction rates as a function of temperature.  The solid (red) line is for $\lambda_{\nu_en}$, the dashed (green) line is for $\lambda_{e^+n}$, the dotted (blue) line is for $\lambda_{n_{\rm decay}}$, the small-dashed(pink) line is for $\lambda_{p\bar\nu_e}$, the dotted-dashed (cyan) line is for $\lambda_{e^-p}$, and the black dotted-spaced line is for  $\lambda_{pe^-\bar\nu_e}$.  All lepton chemical potentials are set to zero here.}
\label{figure1}
\end{figure}

This figure shows that at a lower temperature ($T\ll\delta m_{np}$) the electron capture rate $\lambda_{e^-p}$ and the three-body rate $\lambda_{pe^-\bar\nu_e}$ track each other closely, differing by a factor of order unity.  This is readily explained as follows.  First note that the integrands in the phase space factors in Eq.~(\ref{genep}) and Eq.~(\ref{revrate}) are identical.  Although the former phase space factor is proportional to $1-S_{\nu_e}$ and the latter to $S_{\bar\nu_e}$, when the $\nu_e$ and $\bar\nu_e$ energies in these distributions are expressed in terms of electron energy,  $E_{\nu_e}=E_e - \delta m_{np}$ and $E_{\bar\nu_e}=\delta m_{np}-E_e$, respectively, we see that $1-S_{\nu_e}=S_{\bar\nu_e}$.  Second, though the limits of integration for these phase space factors differ, we note that when $T\ll\delta m_{np}$ the upper limits are effectively the same.  Only the lower limit is different in the two cases, $\approx 1.3$\,MeV in the former and $\approx 0.511$\,MeV in the latter. 

As the temperature decreases, free neutron decay becomes the dominant weak reaction.  This remains the case through the epoch when strong and electromagnetic nuclear reactions freeze out of equilibrium (\lq\lq nucleosynthesis\rq\rq ), when nearly all free neutrons are incorporated into alpha particles.

\section{Coulomb Correction to the Weak Interaction Rates}

For weak interaction processes that have an electron and a proton in either the initial or final state, the Coulomb interaction must be taken into account.  In fact, the phase space factors presented above are derived by using plane wave functions for the entrance and exit channel leptons, but then \lq\lq correcting\rq\rq\ where Coulomb waves should be used instead by multiplying the appropriate phase space integrals by the Fermi factor $F(Z, E_e)$.

The Coulomb potential is attractive in the $e^-/p$ channel.  This has the effect of {\it increasing} the electron probability amplitude at the nucleus (proton) and, in turn, this will always {\it increase} the affected phase space factors.  In other words, the phase space factors in the expressions for the rates for both the forward and reverse processes in Eq.~(\ref{nuenn}) and Eq.~(\ref{ndecayy}) will be increased over a case where only plane waves are used.

\subsection{Previous Corrections to the BBN Weak Reaction Rates}
%wagoner, Discus, T&L, Esposito
The Coulomb correction to the weak rates in BBN was first employed by Wagoner \cite{wag73} in an early version of the BBN code.  Wagoner took a representative value of the correction from around the time of weak freeze out, when the weak rates have the largest effect on the $n/p$ ratio, and used this to scale the neutron lifetime, $\tau_n$.   This had the effect of increasing the effective neutron lifetime over what had been chosen for $\tau_n$ by about $2\%$.  This effectively slowed down all weak interaction rates by $2\%$ because the weak rates are all normalized by the neutron lifetime.  Wagoner's correction over-estimated the Fermi factor, giving an erroneous increase in helium mass-fraction, $Y_p$, of about $0.5\%$.  This over production was largely a result of \lq\lq correcting\rq\rq\ weak reactions that should not have been corrected, {\it e.g.}, $n+e^+ \rightleftharpoons p+ \bar\nu_e$.  Wagoner did not include any radiative corrections.

Dicus {\it et. al} \cite{dicus} were the first to add an energy dependent Coulomb correction to the BBN calculation, along with the zero-temperature radiative corrections and finite-temperature radiative corrections.  They approximated the Coulomb correction using the non-relativistic form for the Fermi factor
\begin{equation}
F_+(\beta)\simeq \frac{2\pi\alpha/\beta}{1-e^{-2\pi\alpha/\beta}}.
\label{nonrel}
\end{equation}
where $\beta=v/c$ is the electron velocity. It was pointed out in Ref.~\cite{kandk} that Ref.~\cite{dicus}, like Wagoner, Coulomb-corrected the rates that should not have had a Fermi factor in their phase space integrands.  

Ref.~\cite{dicus} derived the zero-temperature radiative corrections for a point nucleon, finding that all the weak rate integrands should be multiplied by
\begin{equation}
1+ {{\alpha}\over{2\pi}} C(\beta,y),
\end{equation}
where $\beta$ is again the electron velocity and $y$ and $\epsilon$ are the neutrino energy and electron energy divided by the electron mass, respectively, and
\begin{eqnarray}
C(\beta,y)& \cong & 40+4(U-1)(y/3\epsilon -3/2+\ln 2y)\\
&& +U(2(1+\beta^2)+y^2/6\epsilon^2-4\beta U]\nonumber \\
&& - 4[2+11\beta+25\beta^2+25\beta^3+30\beta^4\nonumber \\
&&+20\beta^5+8\beta^6)/(1+\beta)^6.
\nonumber
\end{eqnarray}
Here $U$ is defined to be
\begin{equation}
U \equiv \beta^{-1}\tanh^{-1}\beta.
\end{equation}
The corrections in Ref.~\cite{dicus} resulted in a $\sim 0.4\%$ reduction in $Y_p$ from a calculation with the Coulomb effect alone, plus a $\sim 0.2\%$ increase stemming from the zero-temperature radiative corrections.

Ref.~\cite{esposito}  and Ref.~\cite{tandl} appropriately applied the Coulomb correction to only those rates which require one.  Ref.~\cite{esposito} used the Fermi factor approximated at order $\alpha$, while Ref.~\cite{tandl} used a non-relativistic version of the Fermi factor in Eq.~(\ref{nonrel}). They also applied the zero-temperature radiative corrections defined above as well as several other corrections.

\subsection{New Coulomb Correction and Modifications to the BBN code}
%explain our modifications to the code, and explain how our correction is different and our version of the code is different.
In this work we employ a version of the Coulomb correction which can better take into account the potentially relativistic kinematics of initial or final state electrons. We use the Coulomb correction that is discussed in Ref.s~\cite{FFNI, gandm, deshalit}:
\begin{equation}
 G(\pm Z, E_e) \equiv x\,F(\pm Z, E_e),
 \label{g}
 \end{equation}
and we define $x \equiv (E_e^2-(m_ec^2)^2)^{1/2}/E_e$, the ratio of charged lepton (electron here) momentum to energy.  In Eq.~(\ref{g}), $F(\pm Z, E_e)$ is the Fermi factor (or relativistic Coulomb barrier factor) approximated here as
 \begin{equation}
 F(\pm Z,E_e) \approx 2(1+s)(2pR)^{2(s-1)}e^{\pi\omega}\Bigg\vert{{\Gamma(s+i\omega)}\over{\Gamma(2s+1)}}\Bigg\vert.
 \end{equation}
In this expression, the upper signs are for electron emission and capture, the lower signs are for positron emission and capture in the general case for a nucleus of electric charge $Z$ (in our case  $Z=1$), $s=[1-(\alpha Z)^2]^{1/2}$, $\alpha$ is the fine structure constant, $\omega = \pm Z/x$ (\lq\lq+\rq\rq\ for the $e^-$ in our cases), and $R$ is the nuclear radius in electron Compton wavelengths, $R= 2.908\times 10^{-3} A^{1/3} - 2.437A^{-1/3}$ where $A$ is the nuclear mass number and A=1 in our case.  This is the most accurate Coulomb correction that has been employed in a BBN calculation.
 
In order to properly apply these features of the correction, we used a version of the Kawano/Wagoner code where the weak reaction rates have been entirely rewritten.  We will only briefly describe this code here.  A detailed description can be found in Ref.~\cite{ourcode}.  

The original Kawano/Wagoner code calculates the weak rates with a total lumped sum of the $n \rightarrow p$ and $p \rightarrow n$ rates:
\begin{equation}
\lambda_{n} = \lambda_{\nu_e+n\rightarrow p+e^-} + \lambda_{n+e^+\rightarrow p+\bar\nu_e} + \lambda_{n\rightarrow p+e^-+\bar\nu_e}
\label{n-rates}
\end {equation}
\begin{equation}
\lambda_p = \lambda_{p+e^-\rightarrow \nu_e+n} + \lambda_{\bar\nu_e+p\rightarrow n+e^+} + \lambda_{p+e^-+\bar\nu_e\rightarrow n}.
\label{p-rates}
\end{equation}
In our version, we have separated these summed rates to calculate all 6 weak reaction rates individually.  Another key feature of the code is that each rate is modularized, so that any neutrino and antineutrino distribution function and time dependence thereof can be applied.

Because of this modularization we were able to apply a Coulomb correction to only those rates that require one.  In other words, we were also able to include an appropriate relativistic Fermi factor in the integrand of those weak rates. 

Additionally, this is the first time this version of the correction has been applied to the full reaction network in the Kawano/Wagoner code.  This allowed us to study the effect of the correction on all of the light element abundances.  We were also able to study the effect of the correction on nucleosynthesis in the presence of neutrino degeneracy (a lepton number).

\section{Results and Discussion}

We have applied the Coulomb correction described above along with zero temperature radiative corrections in the full Kawano/Wagoner BBN code.  The integrated effect of these corrections can be seen by the changes in the light element abundances.  

The Coulomb correction described above affects BBN abundance yields in a subtle, but interesting way which gives insight into the weak interaction's role in setting the neutron abundance in the early universe.  First, as outlined in the last sections, the key effect of calculating the weak rates with Coulomb waves instead of plane waves is to increase the electron's probability amplitude at the proton.  This means that the rates corresponding to $e^-+p\rightarrow n+\nu_e$, $\nu_e+n\rightarrow p+e^-$, $n\rightarrow p +e^-+\bar\nu_e$, and $p+e^-+\bar\nu_e\rightarrow n$ will all {\it increase} over plane-wave calculated rates.  

This is true, but in the BBN calculation the net effect of adding a Coulomb or radiative correction which {\it increases} the phases space factor for free, vacuum neutron decay is to {\it reduce} the weak matrix element ({\it increase} $\langle ft\rangle$) common to all of the rates of the forward and reverse processes in Eq.~(\ref{nuenn}), Eq.~(\ref{nuebarpp}), and Eq.~(\ref{ndecayy}).  This is because for a given vacuum ($S_e=S_{\bar\nu_e}=0$) neutron lifetime, $\tau_n$, we set $\lambda_{n_{\rm decay}}\vert_{\rm vacuum}=\tau^{-1}$, and an increased phase space factor then implies an increased value for $\langle ft\rangle$.

Therefore, the chief effect of a Coulomb correction-mediated increase in phase space factors is a decrease in the overall strength of the weak interaction.  In turn, a weaker weak interaction would cause a higher temperature for freeze-out from chemical equilibrium and a concomitant increase in the neutron-to-proton ratio emerging from the weak freeze out process.  Since, to first approximation, all neutrons will eventually be incorporated into alpha particles, the phase space factor-enhancing Coulomb correction should give rise to an {\it increase} in the primordial mass fraction $Y_p$.

In broad brush this is indeed what the BBN calculations show.  Our modularized code, which allows us to track the individual weak rates, affords us a deeper insight into what is happening.  Though all rates are renormalized downward by the Coulomb correction, the rates for the particular processes with a Fermi factor in their phase space integrals (Eq.s~(\ref{genep}), (\ref{nueonrate}), (\ref{ndecayrate}), and (\ref{revrate})) are decreased less.  In other words, they are increased relative to the rates in Eq.~(\ref{nuonprate}) and Eq.~(\ref{eonnrate}), $\lambda_{\bar\nu_ep}$ and $\lambda_{e^+n}$, respectively.

Nevertheless, neutron decay has more leverage over the eventual $n/p$ ratio than do the lepton capture processes.  At higher temperatures where the n/p ratio is well approximated as the ratio of the sum of the neutron production rates to the sum of the neutron destruction rates, Eq.~(\ref{ntoppp}), note that the small and comparable fractional relative increases in $\lambda_{e^-p}$ and $\lambda_{\nu_en}$ tend to compensate each other to first order.  This is because $\lambda_{e^-p}$ is in the numerator and $\lambda_{\nu_en}$ is in the denominator in Eq.~(\ref{ntoppp}).  As a consequence, changes in $n/p$ stemming from lepton capture processes are second order in the Coulomb corrections.

Table I presents a comparison of BBN calculations of the $^4$He mass fraction, $Y_p$, and the deuterium abundance relative to hydrogen, D/H, all performed with our code.  Shown in this table are a \lq\lq Baseline,\rq\rq\ standard ($\mu_{\nu_e}=\mu_e=0$) BBN case with no Coulomb corrections ($i.e.$, $F(Z,E_e)=1$), and cases where the same calculations were done but where the Coulomb corrections of Wagoner Ref.~\cite{wag73}, Esposito {\it et al.} Ref.~\cite{esposito}, and Lopez and Turner Ref.~\cite{L&T}, were used.  This table also shows results from the same BBN calculation but using our new relativistic Coulomb correction for cases with and without a radiative correction.  Consistent with the arguments given above, we see that the Coulomb correction prescriptions in all of these cases give a $\sim 1\%$ increase in $Y_p$.  All of the various versions of the Coulomb correction are consistent with each other.  The relativistic Fermi factor used in this work results in a modest $0.04\%$ increase in $Y_p$ over the result in Ref.~\cite{esposito}.

\begin{center}
\begin{table}
\caption{$^4$He mass fraction $Y_p$ and deuterium abundance D/H as calculated with our code for various implementations of Coulomb and radiative corrections as indicated.  The Baseline table entries are the uncorrected values; table entries designated by Wagoner, Esposito, and Lopez and Turner were computed using the correction prescriptions in Ref.\,\cite{wag73}, \cite{esposito}, and \cite{L&T}, respectively, but with our code and with the current world-average neutron lifetime.}
\begin{tabular}{ | l | r | r | r | }
\hline
&$Y_p$&$D/H$&$\Delta Y_p/Y_p$\\
\hline
Baseline&0.239&$2.522\times 10^{-5}$&\\
Wagoner&0.2427&$2.543\times 10^{-5}$&$1.16\%$\\
Esposito&0.2416&$2.537\times 10^{-5}$&$1.09\%$\\
Lopez and Turner& 0.2416&$2.537\times 10^{-5}$&$1.09\%$\\
New Correction&0.2417&$2.543\times 10^{-5}$&$1.13\%$\\
New Correction&0.2422&$2.542\times 10^{-5}$&$1.34\%$\\
with Zero-Temperature&&&\\
Radiative Corrections&&&\\
\hline
\end{tabular}
\end{table}
\end{center}

The modular nature of the weak rates in our BBN code allows us to examine the effects of the Coulomb correction for scenarios in which the lepton numbers residing in the $\nu_e$ and $\bar\nu_e$ seas are nonzero.  This is the first such study of the Coulomb and radiative correction effects in a case with nonzero values for $\mu_{\nu_e}$ and $\mu_{\bar\nu_e}$.  (Here we take $\mu_{\nu_e}+\mu_{\bar\nu_e}=0$, reflecting assumed neutrino chemical equilibrium at high temperatures.)  We define the lepton number residing in these neutrino seas to be
\begin{equation}
L_{\nu_e}= {{n_{\nu_e}-n_{\bar\nu_e}}\over {n_\gamma}},
\end{equation}
where $n_{\nu_e}$ and $n_{\bar\nu_e}$ are the $\nu_e$ and $\bar\nu_e$ proper number densities, respectively, and $n_\gamma= (2\xi(3)/\pi^2)T_\gamma^3$ is the corresponding photon number density with $\xi(3)\approx 1.20206$ the Riemann-Zeta function of argument 3.  The primordial helium abundance plus observationally-and experimentally-determined neutrino flavor oscillation data restrict $|L_{\nu_e}|<0.1$ \cite{abb, wong, dolgov, Kneller:2001cd}.  This upper limit conceivably could rise to $\approx 0.2$ if allowance is made for additional contributions to the energy density in the early universe \cite{Kneller:2001cd, barger}.  Models which attempt to reconcile light-mass sterile neutrinos with BBN and large scale structure plus cosmic microwave background-derived overall neutrino mass closure constraints usually invoke lepton numbers.  But the lepton numbers invoked in these schemes can increase the $^4$He-based upper limit on $|L_{\nu_e}|$ \cite{sfka, cirelli}.  Sterile neutrino dark matter scenarios \cite{Kusenko:2004qc, Petraki:2007gq, Dodelson:1993je, afp, Dolgov:2000ew, Shaposhnikov:2006xi, Kusenko:2006rh, Petraki:2007gq, Petraki:2008ef, Shi:1998fu, Chiu:1977ds, Boyanovsky:2007zz, Boyanovsky:2006it, Boyanovsky:2007ba, Abazajian:2002yz, kishimoto:023524} also can invoke significant lepton numbers.  We therefore consider a range $0\leq L_{\nu_e}\leq 0.3$ as an interesting example.

\begin{figure}
\includegraphics[width=2.5in,angle=270]{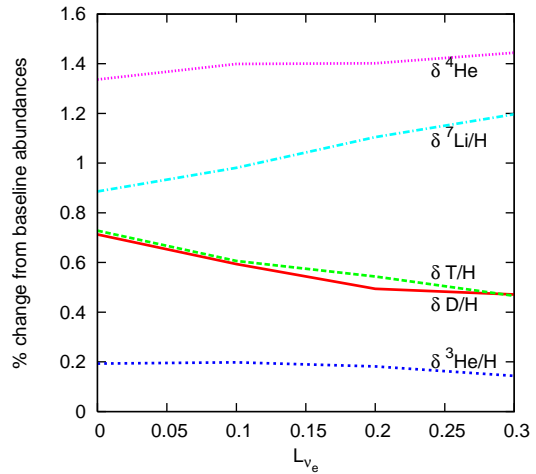}
\caption{Percent change in nuclear abundances from the Baseline values as a function of electron lepton number from when the Coulomb and zero-temperature radiative corrections are included.}
\label{figure2}
\end{figure}

Unlike the neutrino degeneracy parameter, $L_{\nu_e}$ is not a co-moving invariant.  The ratio of the neutrino temperature parameter, $T_\nu$, in $S_{\nu_e}$ and $S_{\bar\nu_e}$ to the temperature of the plasma, $T_\gamma$, evolves in time. This ratio starts out as unity for $T\gg 1$~MeV and, as electrons and positrons annihilate and transfer their entropy preferentially to the photons and plasma, eventually asymptotes to $(4/11)^{1/3}$ at low temperatures.  The lepton numbers and the neutrino degeneracy parameters are related by 
\begin{equation}
L_{\nu_\alpha} = {\left( {{\pi^2}\over{12\zeta\left( 3\right)}} \right)}
{\left( {{T_{\nu}}\over{T_{\gamma}}} \right)}^3 \left[ \eta_{\nu_\alpha}+
\eta_{\nu_\alpha}^3/\pi^2 \right],
\label{leta}
\end{equation}
and at small lepton number $\eta_{\nu_e} \approx 1.46 L_{\nu_e}$.  Here we consider only positive values of $L_{\nu_e}$, $i.e.$, cases with a preponderance of $\nu_e$'s over $\bar\nu_e$'s, as these are the most interesting with respect to $^4$He.

An effect of nonzero $\mu_{\nu_e}$ and $\mu_{\bar\nu_e}$ will be to change the energy weighting in the phase space integrands in Eq.~(\ref{genep}), Eq.~(\ref{nueonrate}), Eq.~(\ref{ndecayrate}), and Eq.~(\ref{revrate}).  We might then expect a concomitant alteration in the effect of the Coulomb correction over the zero lepton number cases for $\lambda_{e^-p}$, $\lambda_{\nu_en},  \lambda_{n_{\rm decay}},$ and $\lambda_{pe^-\bar\nu_e}$, respectively.  In Fig.~\ref{figure2} we show the relative change in BBN abundance yields for $^4$He, $^7$Li, $^3$H, $^2$H, and $^3$He, over the Baseline (no Coulomb or radiative corrections) case as a function of $L_{\nu_e}$.  In the figure the percent change in the $^4$He mass fraction is designated by $\delta^4$He, while the percent change in the abundances for deuterium, tritium, and $^3$He are given by $\delta$D/H, $\delta$T/H, and $\delta^3$He/H, respectively.  First, we note that the overall sense of the Coulomb correction is to increase $Y_p$ and all of the nuclear abundances yields over the Baseline case for the lepton number range examined.  

A higher $L_{\nu_e}$ and the accompanying higher number density $n_{\nu_e}$ and lower number density $n_{\bar\nu_e}$, will have the effect of increasing $\lambda_{\nu_en}$ and $\lambda_{n_{\rm decay}}$ and decreasing $\lambda_{e^-p}$ and $\lambda_{pe^-\bar\nu_e}$ over the zero lepton number case.  This is simply a result of an enhancement or reduction in final state blocking or entrance channel lepton number density as appropriate.  Figure 3 shows that while the $^4$He and $^3$He Coulomb correction abundance yield enhancements are essentially flat with increasing $L_{\nu_e}$, the abundance yield enhancement for $^7$Li increases while the corresponding enhancements for $^3$H and $^2$H decrease with increasing $L_{\nu_e}$.  The trends with $L_{\nu_e}$ of $^7$Li/H, $^3$H/H, and $^2$H/H, versus those for the $^{3,4}$He yields reflect the different times at which these species are formed and the sensitivity of the relevant reaction production mechanisms to the local neutron abundance and temperature.

\section{Conclusion}
We have for the first time implemented a relativistic version of the Coulomb correction in the full reaction network of the Kawano/Wagoner BBN code.  We have used this code to study BBN abundance yields for a range of neutrino chemical potentials.  We find that the fully relativistic Coulomb correction essentially agrees with previous non-relativistic prescriptions, giving only a 0.04$\%$ increase in the $^4$He yield over that in Esposito {\it et al.} \cite{esposito}.  Our calculations show interesting trends in the light element abundance yields with increasing electron lepton number.  The modularization of the individual weak interaction processes in our code has allowed us to gain insights into how the rates for these processes are altered by the Coulomb and radiative corrections and how these alterations affect the neutron-to-proton $n/p$ ratio in the early universe during the BBN epoch.  In particular, we point out that the lepton capture processes produce changes in the $n/p$ ratio which are only second order in the small Coulomb corrections.

As the accuracy of measurements of the CMB and the primordial abundances of the light elements  increase, BBN will give even better constraints on physics in the early universe.  Currently, the  increase in precision gained from including this relativistic version of the correction is probably unnecessary. However, in the future as the measurements for the main parameters affecting BBN, such as the baryon-to-photon ratio and the neutron lifetime, increase in precision, it may be beneficial to include this version of the correction.  

\begin{acknowledgments}

We would like to acknowledge helpful discussions with Kevork Abazajian, Chad Kishimoto, and Michael Smith.  This work was supported in part by NSF Grant No. PHY-06-53626.
\end{acknowledgments}

\bibliography{mybiblio}

\end{document}